\documentclass[11pt]{article}
\usepackage{amsmath}
\usepackage{amssymb}
\usepackage{bbm}
\usepackage{epsfig}
\allowdisplaybreaks[4]
\begin{document}
\title{Tree-Level Vacuum Stability in Multi Higgs Models}
\author{A. Barroso~\footnote{barroso@cii.fc.ul.pt},
P.M. Ferreira (Speaker)~\footnote{ferreira@cii.fc.ul.pt} and 
R. Santos~\footnote{rsantos@cii.fc.ul.pt}\\ 
Centro de F\'{\i}sica Te\'orica e Computacional, Faculdade de Ci\^encias,\\
Universidade de Lisboa, Av. Prof. Gama Pinto, 2, 1649-003 Lisboa, Portugal \\
 \\
 \\
{\em Talk presented at the International Europhysics Conference on High Energy
Physics} \\
 {\em July 21st - 27th 2005,  Lisboa, Portugal}}
\maketitle
\noindent

{\bf Abstract.} 
In the most general model with two Higgs doublets, if a minimum that
preserves the $U(1)_{em}$ symmetry exists, then charge breaking (CB) minima 
cannot occur. The depth of the potential at a stationary point that breaks CB or
CP, relative to the $U(1)_{em}$ preserving minimum, is proportional to the 
squared mass of the charged or pseudoscalar Higgs, respectively.

\vspace{1cm}

In this talk we will review recent results~\cite{pot10,pot14} on symmetry
breaking in two Higgs doublet models (2HDM)~\cite{2hdm}. Namely, the possibility
that minima that break different symmetries can appear simultaneously in the 
potential, and tunneling between them might occur. In ref.~\cite{pot10} we
worked in 2HDM without explicit CP breaking and showed that if there is, at 
tree level, a minimum that preserves the $U(1)_{em}$ and CP symmetries, that 
minimum is the global one. Therefore, the stability of this minimum is 
guaranteed and tunneling to a deeper one that breaks charge conservation or CP 
becomes impossible. In ref.~\cite{pot14} we extended this analysis to the
most general 2HDM and proved that, once a charge-preserving minimum exists, 
any charge breaking (CB) stationary point that might exist lies above the
minimum. Charge conservation is thus assured. 

There are many  ways of writing the 2HDM tree-level potential, for this talk we 
will use the one introduced in ref.~\cite{vel}. With two scalar Higgs doublets 
in the theory, $\Phi_1$ and $\Phi_2$, both having hypercharges $Y=1$~\footnote{
The numbering of the real scalar $\varphi$ fields is chosen for convenience of
writing the mass matrices for the scalar particles.}, 
\begin{equation}
\Phi_1 = \begin{pmatrix} \varphi_1 + i \varphi_2 \\ \varphi_5 + i \varphi_7
\end{pmatrix} \;\; , \;\; \Phi_2 = \begin{pmatrix} \varphi_3 + i \varphi_4 \\
\varphi_6 + i \varphi_8 \end{pmatrix} \;\; ,
\end{equation}
there are four $SU(2)_W \times U(1)_Y$ invariants one can construct with these
fields, namely
\begin{align}
x_1\,\equiv\,|\Phi_1|^2 &= \varphi_1^2 + \varphi_2^2 + \varphi_5^2 + \varphi_7^2
\nonumber \\
x_2\,\equiv\,|\Phi_2|^2 &= \varphi_3^2 + \varphi_4^2 + \varphi_6^2 + \varphi_8^2
\nonumber \\
x_3\,\equiv\,Re(\Phi_1^\dagger\Phi_2) &= \varphi_1 \varphi_3 + \varphi_2
\varphi_4 + \varphi_5\varphi_6 + \varphi_7\varphi_8 \nonumber \\
x_4 \,\equiv\,Im(\Phi_1^\dagger\Phi_2) &= \varphi_1 \varphi_4 -  \varphi_2
\varphi_3 + \varphi_5 \varphi_8 - \varphi_6 \varphi_7 \;\;.
\label{eq:x}
\end{align}
Notice that under a particular CP transformation in this basis ($\Phi_1 
\rightarrow \Phi_1^*\;,\;\Phi_2 \rightarrow \Phi_2^*$) the invariants $x_1$, 
$x_2$ and $x_3$ remain the same but $x_4$ changes sign. The most general 
tree-level potential is thus composed of all the terms linear and quadratic in 
the $x$'s, i.e., 
\begin{align}
V \;\;=& \;\; a_1\, x_1\, + \,a_2\, x_2\, + \,a_3 x_3 \,+\, a_4 x_4 \,+\, b_{11}
\, x_1^2\, +\, b_{22}\, x_2^2\, +\, b_{33}\, x_3^2\, +\, b_{44}\, x_4^2\, +\,
\nonumber \\
 & \;\; b_{12}\, x_1 x_2\, +\, b_{13}\, x_1 x_3\, + b_{14}\,
x_1 x_4\, +\,b_{23}\, x_2 x_3 +\,b_{24}\, x_2 x_4 +\,b_{34}\, x_3 x_4\;\; .
\label{eq:pot}
\end{align}
The $a_i$ parameters have dimensions of mass squared and the $b_{ij}$
parameters are dimensionless. The potential thus written depends on 14 real
parameters but, with a particular choice of basis, one can reduce this
number to 11 independent parameters (see, for instance,~\cite{hab}). The linear
terms in $x_4$ are the ones that explicitly break CP, and if we eliminate them 
we are left with the 10 parameter CP preserving potential that was used in 
ref.~\cite{pot10} (with a judicious choice of basis~\cite{hab} the number of 
independent real parameters of this potential may be reduced to 9). 

It is a well known fact~\cite{2hdm} that the 2HDM potential can only have three
types of minima. One of them is a charge breaking minimum where three of the 
fields, at least one of them carrying electrical charge, have non-vanishing 
vacuum expectation values (vevs). For instance, the fields $\varphi_5$, 
$\varphi_6$ and $\varphi_3$. In the second possible type of minimum only neutral
fields have vevs and there are two possibilities. In the first only two fields
have vevs ($\varphi_5$ and $\varphi_6$, for instance), and we call this the
first ``normal" minimum, $N_1$. In the second case there are three vevs, for the
fields $\varphi_5$, $\varphi_6$ and $\varphi_7$, for example. We call this case
the $N_2$ minimum. Notice that when the model is reduced to the potential that 
explicitly preserves CP, the $N_2$ minimum spontaneously breaks CP. For this 
more general case, however, there is {\em a priori} no physical distinction 
between the two ``normal" minima, both of them preserving charge conservation. 

In references~\cite{pot10} and~\cite{pot14} we developed a method to compute
the value of the tree-level potential at each of these stationary points, and
compare their value. We refer the readers to those publications for details
of the calculations, and proceed to present the results and their consequences. 
For the $N_1$ minimum, the vevs will be $\varphi_5 = v_1$ and $\varphi_6 = v_2$;
for CB, we will have $\varphi_5 = v_1^\prime$, $\varphi_6 = v_2^\prime$ and 
$\varphi_3 = \alpha$ (this last vev is the charged one that breaks charge 
conservation); for $N_2$, $\varphi_5 = v_1^{\prime\prime}$, $\varphi_6 = 
v_2^{\prime\prime}$ and $\varphi_7 = \delta$ (in the case of the CP preserving 
potential, this last vev is the one that spontaneously breaks CP). The 
difference between the values of the potential at a CB and an $N_1$ stationary 
points is given by
\begin{equation}
V_{CB}\;-\;V_{N_1} \;\; = \;\;\frac{1}{2}\,Y^T\,V^\prime \;\; = \;\;
\frac{M^2_{H^\pm}}{2\,v^2} \;\left[ (v^\prime_1\,v_2
\;-\;v^\prime_2\,v_1)^2\; + \; \alpha^2\,v_1^2\right]\;\;\; ,
\label{eq:difp}
\end{equation}
where $M^2_{H^\pm}$ is the value of the squared charged scalar mass at $N_1$.
Then, if $N_1$ is a minimum, we will necessarily have $M^2_{H^\pm} > 
0$ and, given that the quantity in square brackets above is always positive, we
conclude that $V_{CB}\;-\;V_{N_1}\;> \;0$. Then, the CB stationary point is
clearly above the $N_1$ minimum. Furthermore, it is possible to show that under
these circumstances the matrix of CB squared scalar masses is neither positive
nor negative definite. As a result, we reach the conclusion the the CB 
stationary point is a saddle point, and lies above the $N_1$ minimum. 

Results altogether identical are obtained if one compares the $N_2$ and CB 
potentials. From~\cite{pot14} we see that
\begin{equation}
V_{CB}\,-\,V_{N_2} \; = \;
\left(\frac{M^2_{H^\pm}}{2\,v^2}\right)_{N_2} \left[ (v^\prime_1\,
v^{\prime\prime}_2 \,-\,v^\prime_2\,v^{\prime\prime}_1)^2\, + \, \alpha^2\,
({v^{\prime\prime}_1}^2\,+\,\delta^2)\,+\,\delta^2\,{v^\prime}^2_2\right] \;\; ,
\label{eq:difn2}
\end{equation}
where now $\left(M^2_{H^\pm}\right)_{N_2}$ is the squared charged scalar mass
of the $N_2$ stationary point, and $(v^2)_{N_2}\,=\,{v^{\prime\prime}}^2_1\,+\,
{v^{\prime\prime}}^2_2\,+\,\delta^2$. Again, we reach the conclusion that, if 
$N_2$ is a minimum, then $V_{CB}\,-\,V_{N_2}\;> \;0$, and the CB stationary
point lies above the normal minimum, Again, it is possible to demonstrate that
the CB stationary point is a saddle point. The conclusion to take from this 
analysis is that, if a minimum that preserves electric charge conservation 
exists, it is necessarily deeper than any CB stationary point that might
exist in the model. Further, that stationary point is necessarily a saddle
point. There is therefore no possibility whatsoever of tunneling from a 
charge-preserving minimum to a deeper one where charge is broken, and the
masslessness of the photon is thus assured. 
 
What about a comparison between the values of the potential at $N_1$ and $N_2$
stationary points? Unfortunately we cannot reach any definite conclusion about
which of these possible minima is deeper. Following a chain of thought 
altogether identical to the previous cases, one obtains
\begin{equation}
V_{N_2}\;-\;V_{N_1} \;\; = \;\;\frac{1}{2}\left[
\left(\frac{M^2_{H^\pm}}{v^2}\right)_{N_1}\;-\;\left(\frac{M^2_{H^\pm}}{v^2}
\right)_{N_2} \right] \;\left[(v^{\prime\prime}_1\,v_2 \;-\;
v^{\prime\prime}_2\,v_1)^2\; + \; \delta^2\,v_2^2\right]\;\;\; .
\label{eq:difcp}
\end{equation}
Depending on which stationary point has a larger value for the squared charged
mass, then, either $N_1$ or $N_2$ might be deeper. This seems to depend on the
particular values of the parameters of the model, both cases {\em a priori} 
possible. A very interesting thing happens, though, when we restrict ourselves
to the case of the CP preserving potential. In that case the $N_1$ minimum
preserves both electric charge conservation {\em and} CP, and the $N_2$ 
stationary point spontaneously breaks CP. Calling $V_{N_1} = V_N$ and $V_{N_2} =
V_{CP}$, and investigating the mass matrices of the 2HDM model (for instance, 
\cite{pot10}) we obtain a remarkable result,
\begin{equation}
V_{CP}\;-\;V_{N} \;\; = \;\;\frac{M^2_A}{2\,v^2}\;\left[
(v^{\prime\prime}_1\,v_2 \;-\;v^{\prime\prime}_2\,v_1)^2\; + \; \delta^2\,v_2^2
\right]\;\;\; ,
\label{eq:ma}
\end{equation} 
where $M^2_A$ is, as usual, the squared pseudoscalar mass at the normal (i.e., 
charge and CP preserving) minimum. The right-hand side of this equation is thus 
positive and we have, just as in the CB case, $V_{CP}\;-\;V_{N} \;>\;0$. The CP 
stationary point is therefore {\em above} the normal minimum, but in this case 
it is not obvious whether it is also a saddle point. Thus no tunneling to a 
deeper minimum may occur once the potential is at a vacuum that respects both CP
and charge conservation. The tree level vacuum, we have therefore shown, is 
perfectly stable. 

An intriguing aspect of these results is the following: if one observes
equations~\eqref{eq:difp} and~\eqref{eq:difn2}, one sees that the difference in 
the depth of the potential between the normal minimum and the CB stationary 
point is ``controlled" by the charged Higgs squared mass. On the other hand, 
equation~\eqref{eq:ma} shows that the difference in the value of the potential 
between the CP and the normal stationary points is proportional to the 
pseudoscalar squared mass. That is, the depth of the potential at a stationary 
point that breaks a given symmetry, relative to the normal minimum, depends, in 
a very straightforward manner, on the mass of the scalar particle directly 
linked with that symmetry. The absence of charge breaking when normal minima 
exist seems to be related to the non-existence, in the potential, of cubic 
terms in the fields. In fact, analysing the Zee model~\cite{zee} scalar 
potential - this model consists of the 2HDM plus a charged $SU(2)$ singlet 
scalar -, where such terms are present, CB minima deeper than the normal ones 
are discovered~\cite{zeecb}. This is not surprising, since charge - and colour -
breaking is known to occur in supersymmetric theories~\cite{fre}, for which the
scalar potential has, once again, cubic terms in the fields.

\end{document}